\documentstyle[twocolumn]{article}
\newcommand{\shrinka}{\def\baselinestretch{0.88}\large\normalsize}

\newcommand{\shrinkc}{\def\baselinestretch{1.00}\normalsize\footnotesize}

\newcommand{\PSbox}[3]{\mbox{\rule{0in}{#3}\includegraphics{#1}\hspace{#2}}}

\makeatletter
\def\@normalsize{\@setsize\normalsize{10pt}\xpt\@xpt
\abovedisplayskip 10pt plus2pt minus5pt\belowdisplayskip 
\abovedisplayskip \abovedisplayshortskip \z@ plus3pt\belowdisplayshortskip 6pt plus3pt minus3pt\let\@listi\@listI}
\def\subsize{\@setsize\subsize{12pt}\xipt\@xipt}
\def\section{\@startsection {section}{1}{\z@}{1.0ex plus 1ex minus .2ex}{.2ex plus .2ex}{\large\bf}}
\def\subsection{\@startsection {subsection}{2}{\z@}{.8ex plus 1ex} {.2ex plus .2ex}{\subsize\bf}}
\makeatother
\begin{document}
\date{}
\title{\Large\bf Dynamic Simulation of Splashing Fluids }

\author{ James F. O'Brien and Jessica K. Hodgins\\
\normalsize College of Computing \\
\normalsize Georgia Institute of Technology \\
\normalsize Atlanta, GA 30332-0280\\
\normalsize [obrienj $\vert$ jkh]@cc.gatech.edu\\
}
\maketitle

\thispagestyle{empty}
{\em
\setlength{\parindent}{0pc}
{\bf Abstract: }
In this paper we describe a method for 
modeling the dynamic behavior of splashing fluids. The model 
simulates the behavior of a fluid when objects impact 
or float on its surface. The forces generated 
by the objects create 
waves and splashes on the surface of the fluid. To demonstrate the 
realism and limitations of the model, images 
from a computer-generated animation are presented and 
compared with video frames of actual splashes occuring under
similar initial conditions.
}
\setlength{\parindent}{1pc}

\begin{figure}[b]
\vskip -0.1in
\rule{\columnwidth}{.1mm}
{\footnotesize
From the proceedings of  
{\em Computer~Animation~'95}, pages~198--205.
Held April~19--21,~1995, in Geneva, Switzerland.
}
\end{figure}

\section{Introduction} \label{sec-intro}
The world is filled with natural phenomena that are 
remarkable in their form and movements: 
a tree blowing in the wind, lightning arcing across the sky and the 
simple elegance of ripples in a pool of water. We, along with others in 
the computer graphics community,  are attempting to 
emulate nature's splendor in our computer animations.
Our approach to this problem is to add realism to computer-generated images 
and animations through the use of models and simulations that approximate 
the physical laws of nature.

While many natural phenomena appear simple,
the true complexity of these behaviors is 
considerable. For example, a rock falling into a pool of 
water involves complex interactions between the rock 
and the water. The rock strikes the pool's surface, displacing 
water and creating a disturbance that travels outward 
from the impact as it forms waves on the surface. 
Fluid that is displaced with sufficient force breaks free 
of the pool and flies through the air as spray. Cohesion 
between water molecules causes the formation 
of sheets of water, and air forced beneath the surface 
generates foam and bubbles.

Computer animations have been produced 
by skilled animators manipulating models of a 
computer-generated world using their intuition about the real 
or imaginary physical laws of the animated universe. 
To provide the animator with higher level tools for producing realistic 
computer animations of natural phenomena, researchers have 
investigated the use of dynamically based simulations. By 
modeling the physical effects of natural forces, the modeled
systems can display realistic behavior without 
explicit specification of  the movement for each 
object. Unfortunately, physically correct simulations of 
complex natural phenomena are often computationally 
expensive, requiring many hours of computation time to 
generate only a few seconds of motion.

\begin{figure}[tb]  
\begin{center}
\PSbox{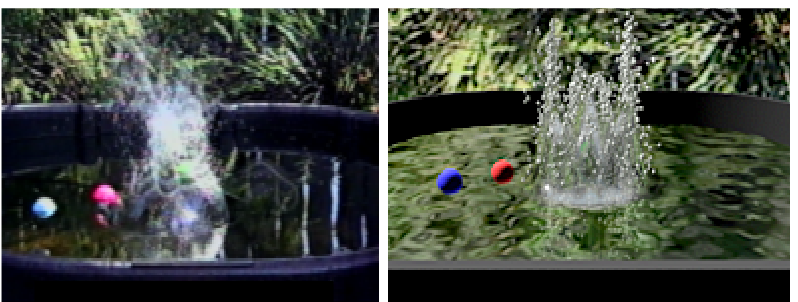} {3.15in}{1.1in}
\end{center}
\vskip -0.2in
\begin{caption}
{
Comparison between a video image of a physical splash and a 
synthetic image generated using dynamic 
models of the fluid and the impacting object.
} \label{fig:lil-splash}
\end{caption}
\end{figure}

In the following sections, we describe a method for 
modeling the dynamic behavior of a fluid. The 
simulation method allows animation of impacts to the surface of the 
fluid, splashes, and the waves 
that arise as a result of an impact. The simulation also 
models the behavior and effects of objects floating on the 
surface of a fluid. Because our goal is to provide a 
method suitable for computer animation, we have 
compromised accuracy for reduced computation time. Nonetheless, 
these methods model many of the visual effects seen in 
the real world, as demonstrated by the images in 
Figures~\ref{fig:lil-splash} and~\ref{fig:comp-spl}.

\section{Background}  \label{sec-bg}
Fluids have been simulated using a number of 
different techniques. One of the most accurate  
is to solve the 3D Navier-Stokes equations describing 
the fluid system \cite{key:knig}. This approach requires 
dividing the space that the fluid occupies into a lattice of cells 
and computing the behavior of the fluid as it moves 
through the cells. Although this approach is well suited for 
scientific simulations requiring a high level of detail, the 
lower bound of the computational cost increases with 
the cube of the model's resolution, making it impractical 
to use this method for interactive computer animation.

Particle systems have been used to model waterfalls and 
other forms of falling water \cite{key:sims}. The dynamics 
equations used in these simulations model particles falling 
under the influence of gravity without modeling 
interactions among the particles. Phenomena such as 
spray or loosely packed water droplets are easily modeled 
by this technique.
Systems that allow particle interactions have been used to
model fountains of fluid and amorphous solids \cite{key:mill}.  While
a large volume could potentially be modeled with interacting particles,
the number of particles needed to fill a volume will grow with the cube of the 
resolution of the model, and despite the simplified equations for the 
dynamics this method can be computationally expensive 
for a large volume of particles.

Fournier and Reeves \cite{key:four} and Peachey \cite{key:peac}
implemented techniques that model the movement 
of explicit waves over the surface of a fluid. They used 
functions that describe the shape of a wave to displace a 
model representing the free surface of the fluid. Spray 
from the top of the wave crests is modeled using a 
particle system. Although these methods work well for modeling 
the propagation of wave-trains, it is more difficult to apply them
to simulations where no explicit set of wave-fronts 
exist, and this approach does not specify
how wave-trains should be created as a result 
of an impact to the surface.

An efficient approach to the problem of simulating solid 
volumes of water is to use a system that models a 
simplified subset of the fluid dynamics. Kass and Miller \cite{key:kass}
used this approach and represented the surface with a height 
field, modeling the flow between adjacent 
columns of fluid. With this method, surface artifacts, such as 
waves, do not need to be explicitly specified
because they arise naturally from the physical 
conditions occurring within the system. The model can be used 
for large volumes of fluid without incurring huge computational 
costs as the size of the height field is proportional 
to the square, rather than the cube, of the resolution.
Chen and Lobo implemented a similar system that uses
a simplification of the 3D  Navier-Stokes Equations to model
fluid behavior in the presence of moving obstacles~\cite{key:chen}.

Although these systems provide a good set of tools for 
modeling the behavior of waves and falling water,  
we would like to model the interaction of water with other objects. 
Our method extends the approach described by Kass 
and Miller to allow a wider range of behaviors such as 
the splashes caused by impacts and the behavior of floating objects.

\section{Simulation Model}  \label{sec-simmod}

\begin{figure} [tb]
\shrinkc
\begin{center}
\PSbox{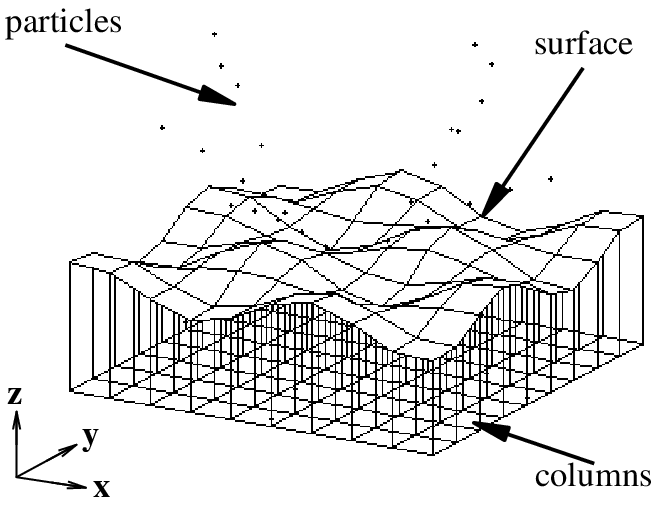} {2.75in}{2.00in}
\end{center}
\vskip -0.2in
\begin{caption} 
{ 
The fluid model is a three part system containing 
subsystems for volume, surface and spray. These 
subsystems interact to model the dynamics of the fluid 
body. 
} \label{fig:diagram}
\end{caption} 
\end{figure} 

To simulate the surface behavior of a body of fluid, 
we use a three-part system where each subsystem 
corresponds to a physical area of the fluid body: the main volume, 
the free surface of the fluid, and disconnected 
components of the fluid (spray). Each subsystem models 
the effects of one aspect of the overall behavior of the 
fluid. Taken together, the subsystems, along with the 
interfaces between them, form the basis for the model (Figure~\ref{fig:diagram}).

\subsection{Volume Model} \label{sec-simmod-vol}

\begin{figure} [tb]
\shrinkc
\begin{center}
\PSbox{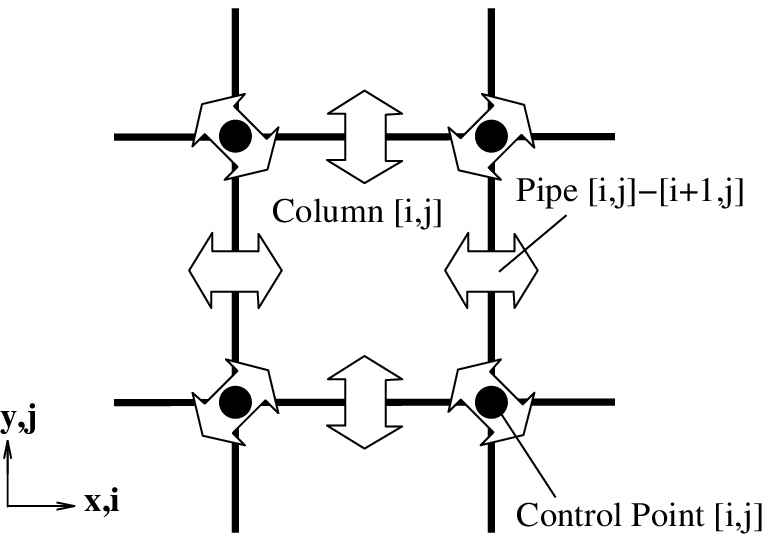} {3.00in}{2.25in}
\end{center}
\vskip -0.2in
\begin{caption} 
{ 
Each vertical column is connected to its eight 
neighbors through a set of directional horizontal pipes. The control 
points for the surface are sampled between the columns 
so that the lines of the surface grid separate adjacent 
columns.
} \label{fig:columns}
\end{caption} 
\end{figure} 

To model the volume that makes up the main 
body of the fluid, we use a formulation that divides the 
body into a rectilinear grid of connected columns. The 
model assumes vertical isotropy within the system and 
all fluid properties within a column are assumed to be 
constant. Flow between these columns occurs through a 
set of virtual pipes that connect adjacent columns. Pipes 
connect columns along the diagonals as well as along the 
axis-aligned  directions (Figure~\ref{fig:columns}).

The equations to determine flow in the pipes are derived 
from the physical laws for hydrostatic pressure. The
static pressure of a column in the grid at position $[i,j]$, 
$H_{ij}$, is 
\begin{equation}
      H_{ij} = h_{ij} \rho g + p_0  \label{eq:head}
\end{equation} 
where $\rho$ is the density of the fluid, 
$g$ is the acceleration due to gravity, 
$p_0$ is the atmospheric pressure in the system,
and 
$h_{ij}$ is the height of the column at $[i,j]$. 
The height of a column is related to the volume of that column, $V_{ij}$,
by 
\begin{equation}
  h_{ij} = { V_{ij} \over d_xd_y }
\end{equation}
where we approximate the cross-sectional column area 
with the product of the nominal distance between mesh 
grid points in the $x$ and $y$ directions, $d_{[x|y]}$. 
When used to determine the flow, equation~(\ref{eq:head}) is 
an approximation because it relies on the assumption that the fluid 
is not moving rapidly \cite{key:muns}. 

In addition to the pressure due to gravity, we model external forces 
to simulate the impact of objects on the surface of the 
fluid. These forces are described in Section~\ref{sec-extern}, but 
for now we include an additional term, the external 
pressure $E_{ij}$, and rewrite equation~(\ref{eq:head}) for total pressure 
$P_{ij}$: 
\begin{equation}
     P_{ij} = h_{ij} \rho g + p_0 + E_{ij} .
\end{equation}

As previously mentioned, each column is connected to 
its eight neighbors by a set of virtual pipes. A flow 
velocity is maintained for each pipe in the system and at 
each time step a force accelerates the flow of fluid in the 
pipe. The force is determined by measuring the pressure 
differential, $\Delta P _{ij \rightarrow kl}$, across the pipe between column 
$[i,j]$ and one of its eight neighbors, $[k,l] \in \eta_{ij}$. Substituting into 
the elementary equation $F=ma$, we can derive the 
acceleration of the fluid in a pipe, $a_{ij \rightarrow kl}$:
\begin{equation}
     a_{ij \rightarrow kl} = {{c( \Delta P _{ij \rightarrow kl}) } \over {m}} \label{eq:acc1}
\end{equation}
where $c$ is the cross-sectional area of the pipe and $m$ is 
the mass of the fluid in the pipe. Expressing the mass of 
the fluid in a pipe of length $l$, as $m = \rho cl$, and 
assuming atmospheric pressure to be constant, equation~(\ref{eq:acc1}) 
can be rewritten:
\begin{equation}
  a_{ij \rightarrow kl} = {  \rho g ( h _{ij} - h _{kl} ) + E_{ij} - E_{kl}  \over \rho l } .
\end{equation}

Assuming that the acceleration is constant over some 
time period, $\Delta t$, the flow in the pipe, $Q_{ij \rightarrow kl}$, is
\begin{equation}
 Q^{t+\Delta t}_{ij \rightarrow kl} = Q^{t}_{ij \rightarrow kl} + \Delta t ( ca_{ij \rightarrow kl} )
\end{equation}
and the net volume change of a column during a time 
interval is
\begin{equation}
 \Delta V _{ij} = \Delta t \sum _{kl \in \eta_{ij}} \left[ { Q^{t+\Delta t}_{ij \rightarrow kl}+Q^{t}_{ij \rightarrow kl} \over 2 } \right] . 
\end{equation}

In order to conserve volume within the system, some 
constraints must be added to the system. Because the 
fluid is incompressible, the flow at one end of a pipe 
must be equal in magnitude with opposite sign to the 
flow at the other end:
\begin{equation}
 	\forall ij: \forall kl \in \eta_{ij}  : Q_{ij \rightarrow kl} = -Q_{kl \rightarrow ij}. 
\end{equation}
A negative volume does not make physical sense, and 
the following inequality ensures that the volume in all columns
remain positive:
\begin{equation}
   V^{t + \Delta t}_{ij} \ge 0 \iff V^{t}_{ij} \ge - \Delta V^{t}_{ij}  .   \label{eq:neg_vol}
\end{equation}
To keep equation~(\ref{eq:neg_vol}) true at the end of each integration 
step, the volume of each column is tested. 
If a column has a negative volume, then 
all pipes that are removing
fluid from that column are scaled back.  This procedure is repeated until all
columns in the grid have a positive volume.

The columns on the outer edges of the grid have pipes 
that lead out of the area where columns are defined. 
Because no columns exist at the other end of the virtual 
pipe, flow conditions must be specified to model boundary 
conditions. To model a barrier, such as a wall, the 
flow is set to zero. Phenomena such as fluid sources or 
sinks can be created by setting the flow to a positive 
or negative constant.

\subsection{Surface Model} \label{sec-simmod-surf}

The surface subsystem allows external objects to interact 
with the fluid system. Objects that collide with or 
float on the surface exert forces on the surface model. 
These forces are propagated as external pressure to the 
volume subsystem. The vertical positioning of the surface 
elements is determined by the volume of the columns.

The model for the surface subsystem is a rectilinear grid 
of control points that define a mesh. The control points 
of the mesh map onto the column grid of the volume 
model so that the control points are sampled between adjacent 
columns (Figure~\ref{fig:columns}). 
The vertical position of a grid point, 
$z_{ij}$, is determined by averaging the 
height of the four columns surrounding the grid point:
\begin{equation}
  z_{ij} = { h_{i,j} + h_{i,j+1}  + h_{i+1,j} + h_{i+1,j+1} \over 4} .
\end{equation}
Forces applied 
to each control point are reformulated as an external 
pressure, $E_{ij}$, applied to the columns contacted by that 
control point:
\begin{equation}
  E_{ij} = - { f_e \over 4 d_x d_y }
\end{equation}
where $f_e$ is the force applied to the control point. A 
downward force (negative $z$ direction) will result in an 
increase in external pressure, hence the negative sign. 
Grid points contact four columns and the external pressure 
is evenly distributed across these columns.

\subsection{Spray Model}  \label{sec-simmod-spray}

To model droplets that are disconnected from the 
main body of the fluid, we implemented a particle 
system using techniques similar to those described by 
Reeves \cite{key:reev} and Sims \cite{key:sims}. 
Particles are created when the 
upward velocity of a portion of the surface exceeds a 
threshold\footnote{The value for the threshold can be adjusted to modify the 
appearance of the splash. Lower threshold values will cause 
more particles to be generated over a larger area. The values 
used to generate the images in this paper were between $2.0$ and $2.5 m/s$.}. 
Because the vertical position of a column 
is determined by the volume of fluid in the column, the 
column's vertical velocity can be
determined by the rate of change in volume (the net
flow into the column from its neighbors):
\begin{equation}
  \dot{h}_{ij} = \sum _{kl \in \eta_{ij}} { Q_{kl \rightarrow ij} \over d_x d_y }. 
\end{equation}
The vertical velocity at the surface is
\begin{equation}
  \dot{z}_{ij} = {   \dot{h}_{i,j}   + \dot{h}_{i,j+1}  
                   + \dot{h}_{i+1,j} + \dot{h}_{i+1,j+1} \over 4}
\end{equation}
and the horizontal velocity of the fluid at the surface is
\begin{equation}
  \dot{x}_{ij} ={ Q_{i,j  \rightarrow i+1,j} + Q_{i,j+1 \rightarrow i+1,j+1} \over 2}
\end{equation}
\begin{equation}
  \dot{y}_{ij} ={ Q_{i,j  \rightarrow i,j+1} + Q_{i+1,j \rightarrow i+1,j+1} \over 2}.
\end{equation}

\begin{figure} [tb]
\shrinkc
\begin{center}
\PSbox{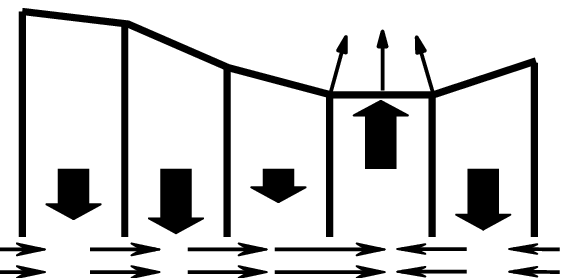} {2.25in}{1.125in}
\end{center}
\vskip -0.2in
\begin{caption} 
{ 
Initial velocity of ejected particles. The horizontal 
arrows at bottom indicate direction of flow between columns, 
the thick vertical arrows indicate upward velocity in 
columns and the thin arrows at top indicate the velocity vectors 
for ejected particles.
} \label{fig:flow}
\end{caption} 
\end{figure}

When an area of the surface has an upward velocity 
greater than the threshold, particles are distributed uniformly over 
that area and the initial velocities for the particles are interpolated from
the surface velocities (Figure~\ref{fig:flow}).

To conserve the total volume in the system, the volume 
of each particle is subtracted from the column from 
which it was created.
Once created, the particles fall under the 
influence of gravity and do not interact with each other.

Particles are removed from the system when they fall 
back onto the surface, and the volume of each particle is 
added to the column that absorbed it. Particles that fly 
away from the fluid body are destroyed when they 
encounter a ground plane or strike some other object.

\section{\bf External Objects}    \label{sec-extern}
When an object collides with the fluid surface, we 
model the motion of the fluid and of the object by computing
the forces resulting from the collision and applying 
them to the fluid surface and the object. The 
behavior of an object with mass $m$, at a location $s_0$, moving
at a velocity $\dot {s}$, is described by:
\begin{equation}
  s = s_0 +\dot {s} t + {1 \over 2} \left( g + { f_{o} \over m } \Delta t^2 \right)
\label{eq:fallthing}
\end{equation}
provided that the force, $f_{o}$, and the acceleration, $g$, 
remain constant over the time interval.

The force acting on the fluid is equal in magnitude with 
opposite sign to the force acting on the object ($f_o = -f_e$). If we 
assume that the object remains in contact with the surface 
and does not bounce off, then the location of the 
mass is determined by the location of the surface. The 
force acting at the impact could be computed by solving 
the nonlinear system derived from equation~(\ref{eq:fallthing}) and the 
fluid model system. 

The force, $f_o$, is bound in the range $[0, f_{max}]$, where $f_{max}$ is 
the force that would cause the mass to have a final position 
equal to the initial position of the surface:
\begin{equation}
  f_{max} = m \left( {2 (z-s_0-\dot{s} t) \over t^2 } - g \right).
\end{equation}
A search could be done to find the force that would minimize 
the difference between the final position of the 
mass and the surface. 


An alternative solution that exchanges accuracy for 
speed is to select $f_o$ in the range $[0, f_{max}]$ based on a heuristic measure of 
the expected deceleration of the mass as it
strikes the surface, and perform the integration on the 
fluid system using the approximated value. Thus 
objects with a small impact face would have lower values, 
and objects with a large impact face would 
have larger values. We used this method to produce the images
for this paper.

\section{\bf Results and Conclusions} \label{sec-results}
\begin{figure*} [bt]
\shrinka
\begin{center}
\vskip -0.5in
\PSbox{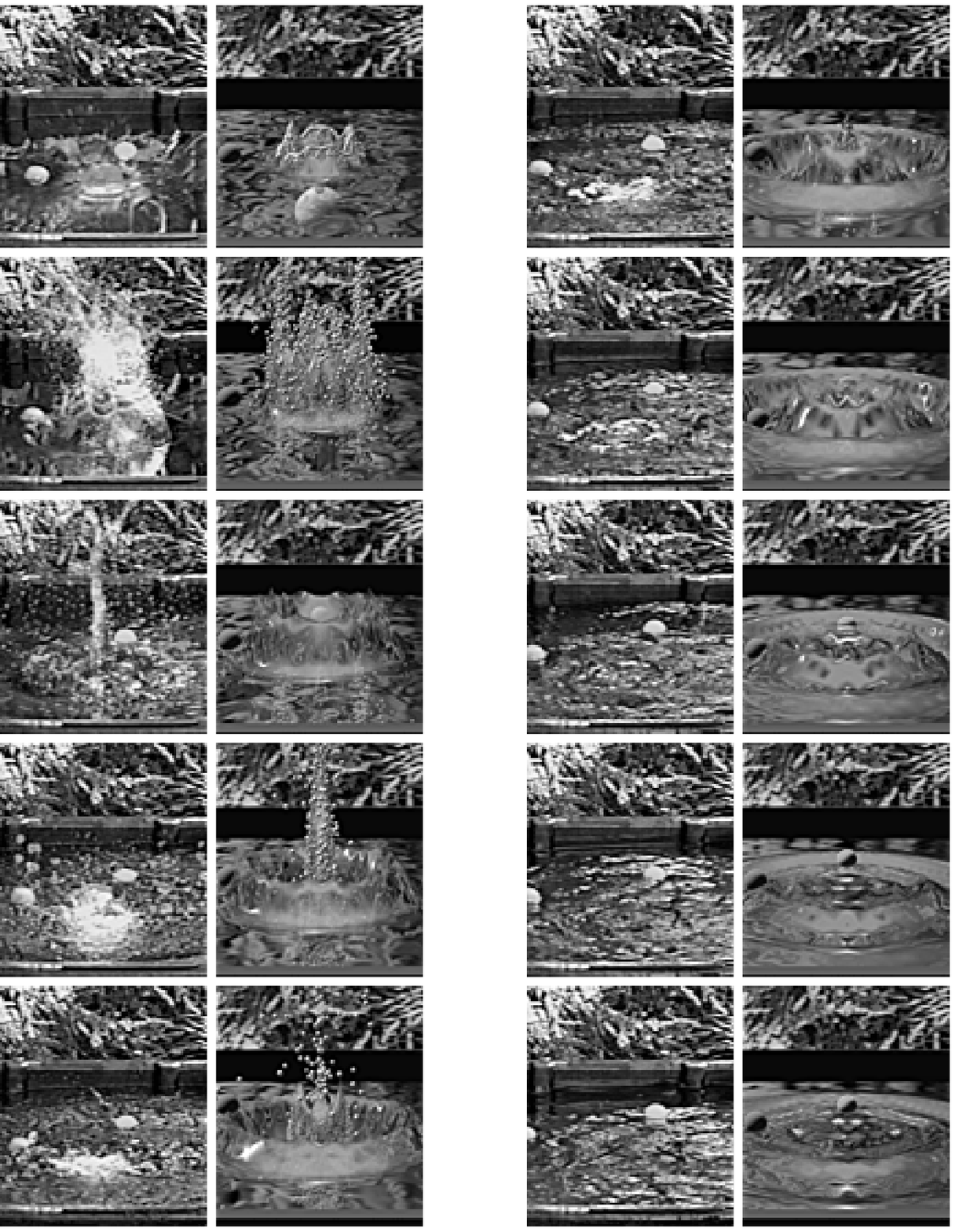} {6in}{7.75in}
\end{center}
\vskip -0.3in
\begin{caption} 
{[See color plate.]
A side-by-side comparison of video of an actual splash
and images from a computer simulation in which we have attempted to 
match the initial conditions to those of the actual splash.  
The images proceed 
top to bottom and left to right and are spaced at 
intervals of $0.233$~seconds.
While the patterns formed in both image sequences are 
qualitatively similar, substantial differences exist between the two. 
For example, the video footage shows sheeting effects that are not present in 
the synthetic images.  The vertical column that appears in the center of the 
actual splash in frame~$3$ does not appear in the simulated splash until frame~$4$.
Finally, the motion of the water in the video footage is more irregular.
} \label{fig:comp-spl}
\end{caption} 
\end{figure*}

In developing our simulation model, we made several 
approximations to simplify the calculations and to 
decrease the required computation time. These simplifications
introduce errors into the final solution. We justify 
this approach by observing that computer animations do not 
generally require the same degree of accuracy as a scientific 
experiment. Furthermore, animation is often used 
to duplicate a known phenomenon, such as a ball falling 
into water, rather than an unknown phenomena, such as 
the air flow within an experimental jet engine. Because 
we are duplicating known phenomena, we can use 
recorded footage or physical intuition to ``tune'' 
the simulation, thus allowing the animator to create
the desired effect while retaining  many of the 
motion characteristics exhibited by water in the real 
world. For example to improve the correspondence shown in Figure~\ref{fig:comp-spl}
we adjusted the time during which the falling object was in contact with the 
water's surface.  Other adjustable parameters include the number of 
particles to be spawned by portions of the surface that have exceeded 
the threshold, the volume of the ejected
particles and the amount of damping used during flow integration.

\begin{figure*} [bt]
\shrinkc
\begin{center}
\PSbox{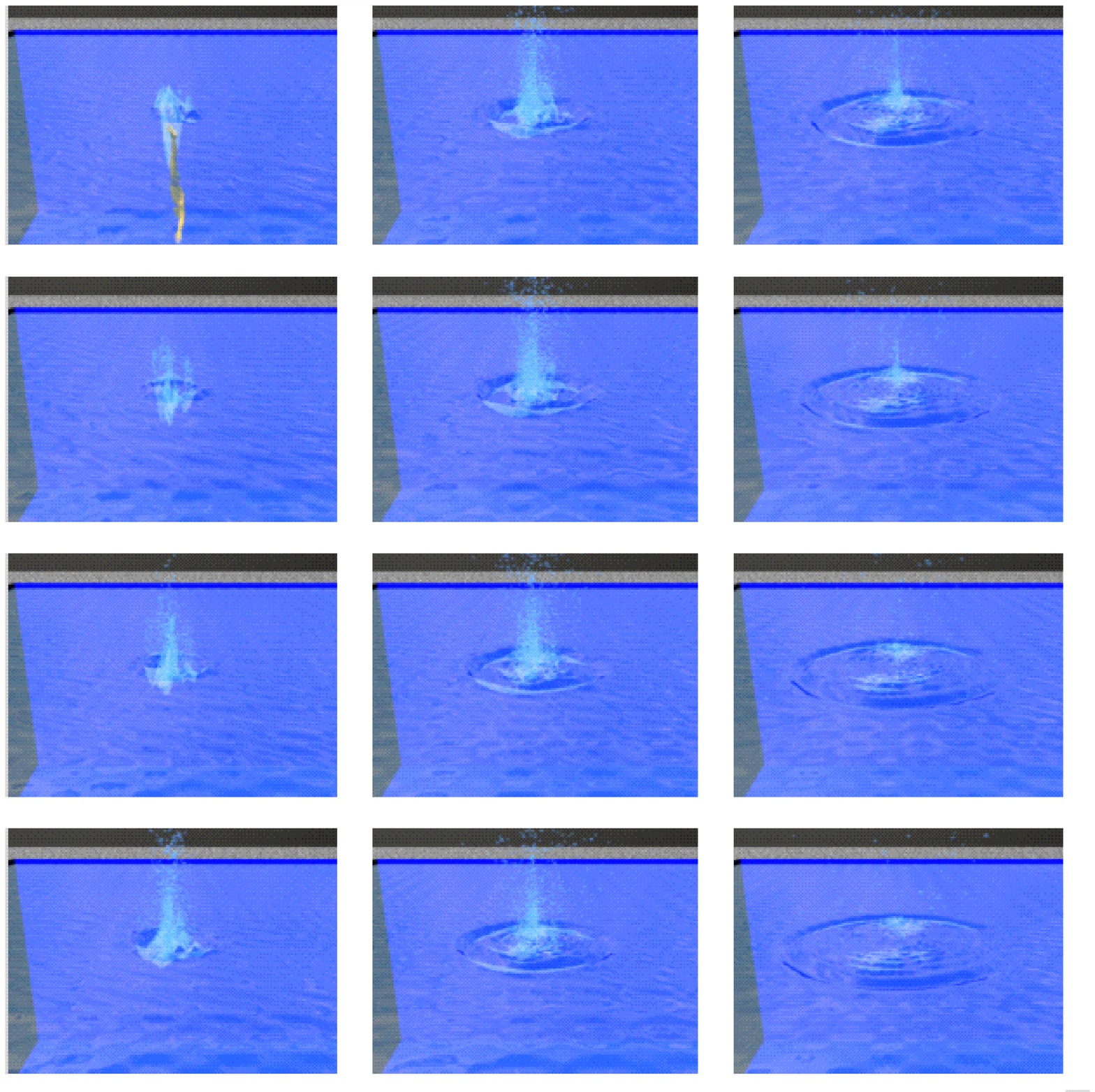} {7.00in}{7.00in}
\end{center}

\begin{caption} 
{ 
Image sequence from a computer-generated animation of a human diver entering the 
water. Frames progress from top to 
bottom and from left to right at intervals of 0.167 seconds. The 
motion of the diver was simulated using rigid body dynamics and control laws for a 
variety of platform dives[10]. 
The images were rendered with Photo-Realistic RenderMan on Silicon Graphics 
workstations. 
}\label{fig:dive-comp}
\end{caption} 
\end{figure*}

The images in Figure~\ref{fig:dive-comp} were rendered from data generated
using the methods described above. These images 
show the splash that results when a physically based 
simulation of a rigid body model of a human diver enters the water after a 
10~meter platform dive \cite{key:woot}. For the purposes of 
computing the intersection of the diver with the pool's 
surface, the human's body was approximated by a set of 
ellipsoids, each representing one of the diver's body 
parts. The pool covers an area of $16 \times 16$~meters 
represented with a $241 \times 241$ grid.

\begin{figure*} [bt]
\shrinkc
\begin{center}
\PSbox{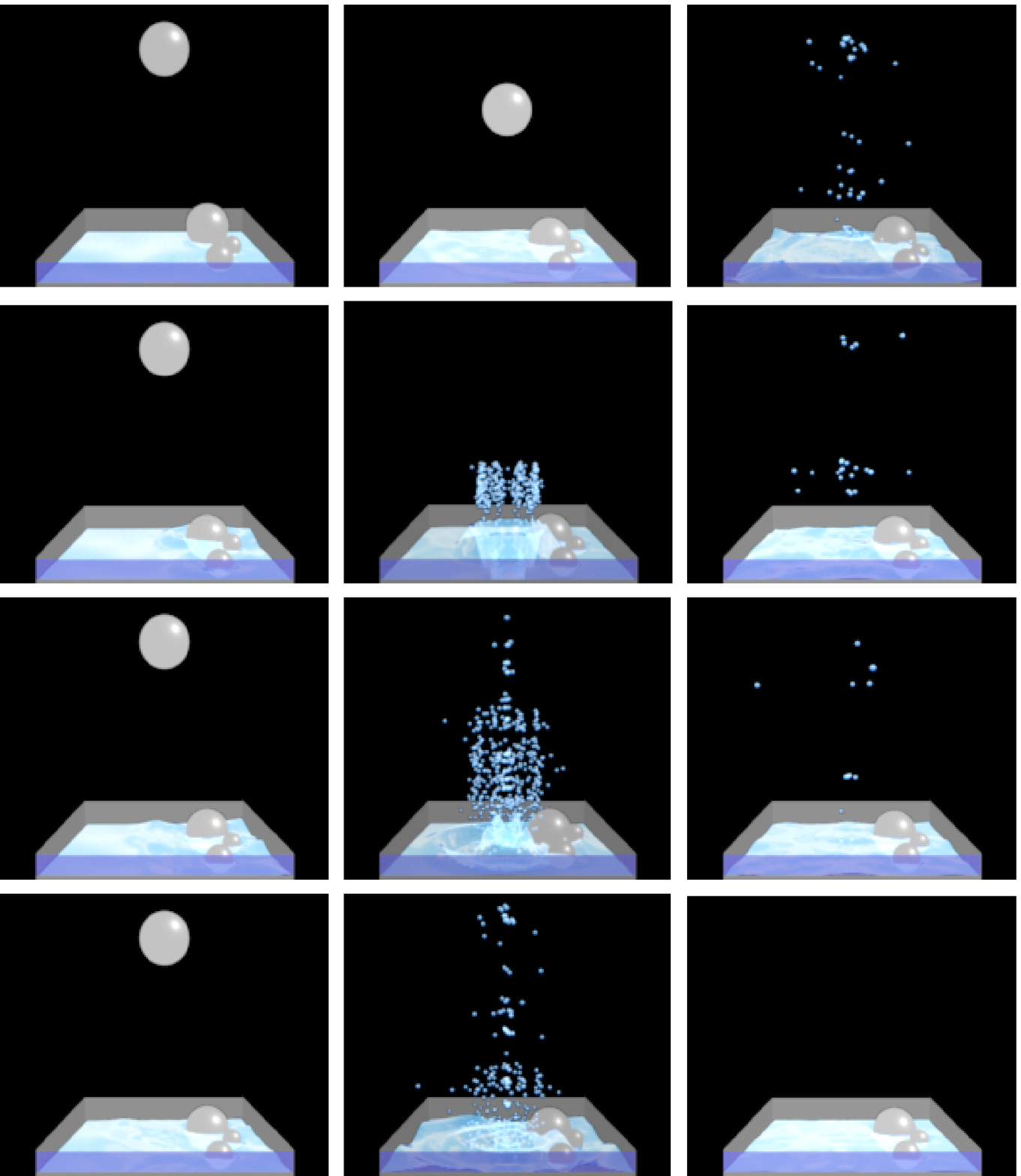} {6.5in}{7.25in}
\end{center}

\begin{caption} 
{ 
Objects floating on a fluid surface. Three balls of equal density 
but different sizes were 
dropped onto the fluid. After the balls settled onto the surface, 
a fourth ball of higher density 
was dropped into the fluid. The floating balls bob on the surface 
as the disturbance caused by the 
heavier ball moves past them.
} \label{fig:balls}
\end{caption} 
\end{figure*}

The system we have implemented runs at interactive 
speeds for simple problems (about one-tenth real-time for a $11 \times 11$ mesh),
but the response time degrades as the resolution 
of the simulation mesh becomes finer and as multiple 
objects interact with the fluid surface. The images in 
Figure~\ref{fig:balls} show three balls floating on a $2$~meter-square 
body of water. They were generated using data from a 
simulation on a $61 \times 61$ grid which  took about 3 minutes 
actual time to complete $2$ seconds of simulation time. 
The simulation for the images in Figure~\ref{fig:dive-comp} took  
longer, a little over an hour for $20$ seconds of 
motion, due to the higher resolution of the grid and the 
larger number of objects. These timings were performed 
on an SGI $Indigo^2$ with an R4000 processor running at 
100mHz. Simulation time 
might be improved by using a non-uniform grid that 
would have a higher resolution in areas where impacts 
would occur and lower resolution elsewhere.

The full spectrum of behaviors exhibited by fluids is 
very complex, and the model we have described allows 
the relatively fast simulation of a subset of these behaviors including 
waves, impacts, splashes, floating objects and other 
behaviors that fit within the assumption
of vertical isotropy. Phenomena that 
occur primarily due to vertical effects  within the fluid, such as turbulence 
out of the horizontal plane,
are beyond the scope of this model. Extending the 
model to include additional horizontal effects or phenomena 
such as flowing streams should be possible as 
would extensions to add foam or bubbles. 

The visual appearance of the model could be improved
by using a more sophisticated spray model that took into account the 
cohesion between drops of water such as the one described in \cite{key:mill}. 
Figure~\ref{fig:comp-spl} shows a comparison
between an actual splash and a simulation where we attempted to match the 
initial conditions of the actual splash.  Although the simulated motion is
similar to that appearing in the captured video images,
a clear difference can 
be seen in the patterns formed by the ejected water droplets.  
The addition of a more sophisticated spray model would allow 
the formation of sheets in the water, improving the perceived 
realism of the splashes.

It is interesting to note that despite 
the obvious differences between the images of real and simulated splashes, 
the resulting computer animation appears realistic  when viewed alone. 
An investigation as to what minimal set of motion qualities 
are essential for  believability might 
enable further simplification and simulation at real-time speeds. 
Real-time execution would allow the inclusion of a simulation
such as this into an interactive virtual reality system.

\vskip 0.2in

High resolution copies of the images in this paper and 
animated image sequences are available through the 
World Wide Web at the URL 

\begin{center}
``http://www.cc.gatech.edu/gvu/animation/
Animation.html''. 
\end{center}

\section*{Acknowledgments}  \label{sec-ack}
The authors would like to thank Wayne Wooten and 
Debbie Carlson for their help in rendering images
and Walter Patterson 
and Adrian Ferrier for their suggestions on simulation methods.

This project was supported in part by NSF Grant No. 
IRI-9309189 and funding from the Advanced Research 
Projects Agency.

\begin{figure*}[bt]
\shrinka
\begin{center}
\vskip -0.5in
\PSbox{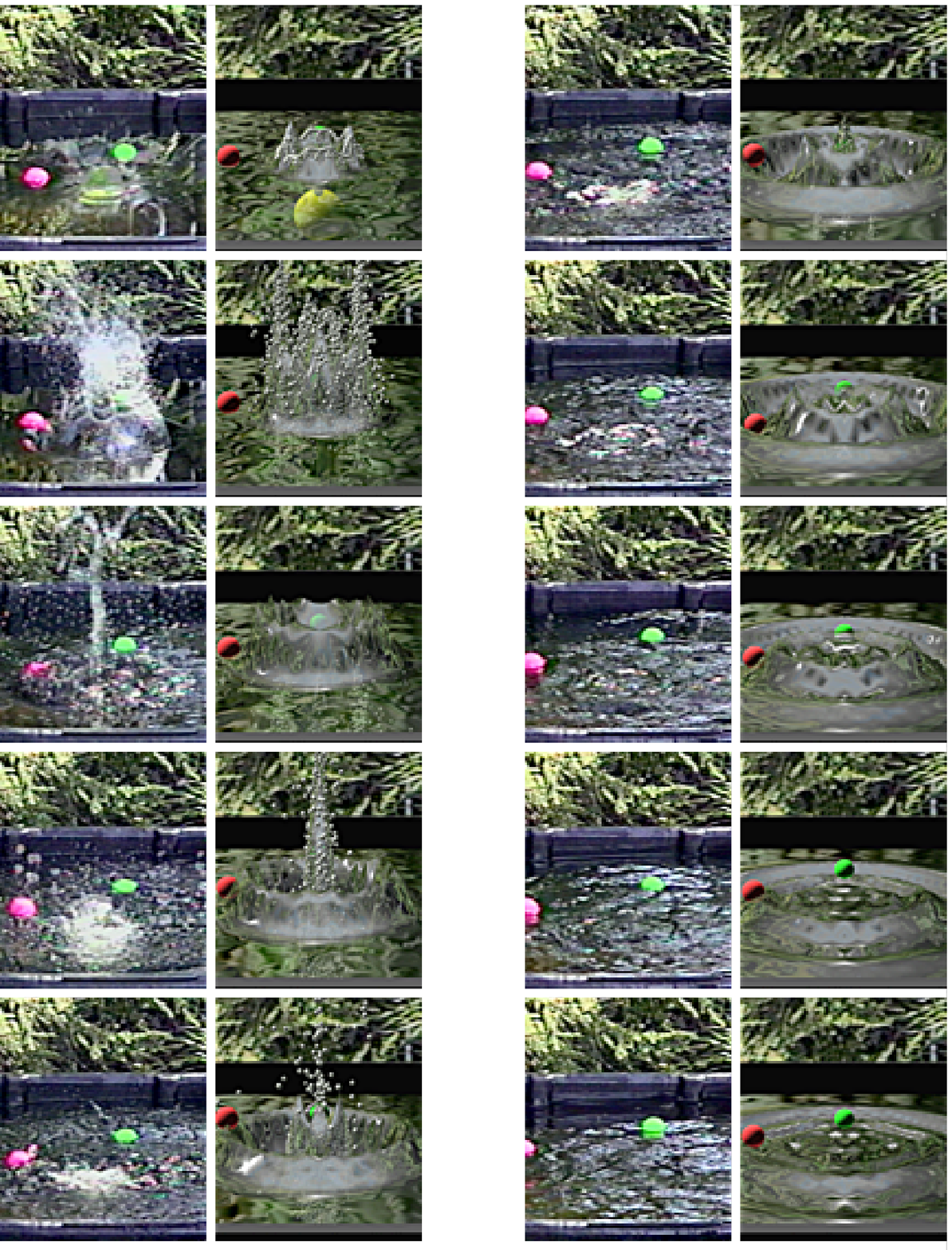} {6in}{7.75in}
\vskip 0.1in
A side-by-side comparison of video of an actual splash
and images from a computer simulation.
\end{center}
\end{figure*} 


\vfill\eject

\end{document}